\documentclass[review, 11pt, sort&compress]{elsarticle}

\usepackage{epsfig}
\usepackage{amssymb}
\usepackage{graphicx}

\usepackage{float}
\floatstyle{plaintop}
\restylefloat{table}
\usepackage[nolists]{endfloat}
\biboptions{square, numbers}

\newcommand{\qq}[1]{{\lq}#1{\rq}}

\newcommand{\marked}[1]{}

\newcommand{\mathnewydd}[1]{#1}
\newcommand{\newydd}[1]{#1}

\journal{Chemical Engineering Science}

\begin{document}
\begin{frontmatter}

\title{Molecular modelling and simulation of the surface tension of real quadrupolar fluids}

\author[TUKL]{Stephan Werth}
\author[ITWM]{Katrin St\"obener}
\author[ITWM]{Peter Klein}
\author[ITWM]{Karl-Heinz K\"ufer}
\author[TUKL]{Martin Horsch\corref{cor}}
\author[TUKL]{Hans Hasse}
\address[TUKL]{University of Kaiserslautern, Laboratory of Engineering Thermodynamics, Erwin-Schr\"odinger Str.\ 44, D-67663 Kaiserslautern, Germany}
\address[ITWM]{Fraunhofer Institute for Industrial Mathematics, Department for Optimization, Fraunhofer-Platz 1, D-67663 Kaiserslautern, Germany}
\cortext[cor]{Corresponding author. E-mail: martin.horsch@mv.uni-kl.de; phone: +49 631 205 3227; fax: +49 631 205 3835.}

\begin{abstract}

Molecular modelling and simulation of the surface tension of fluids with force fields is discussed. 29 real fluids are studied, including nitrogen, oxygen, carbon dioxide, carbon monoxide, fluorine, chlorine, bromine, iodine, ethane, ethylene, acetylene, propyne, propylene, propadiene, carbon disulfide, sulfur hexafluoride, and many refrigerants. The fluids are represented by two-centre Lennard-Jones plus point quadrupole models from the literature. These models were adjusted only to experimental data of the vapour pressure and saturated liquid density so that the results for the surface tension are predictions. The deviations between the predictions and experimental data for the surface tension are of the order of 20 \%. The surface tension is usually overestimated by the models. For further improvements, data on the surface tension can be included in the model development. A suitable strategy for this is multi-criteria optimization based on Pareto sets. This is demonstrated using the model for carbon dioxide as an example. 

\end{abstract}

\end{frontmatter}

\section{Introduction}
\label{sec:introduction}

\noindent
In classical phenomenological thermodynamics following
Gibbs \cite{Gibbs78a}, interfacial properties are considered as excess
contributions which are assigned to a formal dividing surface. In this
way, the surface tension is obtained from the excess free
energy with respect to a
hypothetical system that does not contain an interface, consisting of
the bulk phases in thermodynamic equilibrium only. Theorems that hold
for the bulk properties can be immediately applied to interfacial
thermodynamics, yielding fundamental relations such as the Gibbs
adsorption equation \cite{Gibbs78a,Alberty95}.

In interfacial thermodynamics, the Gibbs dividing surface represents
the highest level of abstraction. Being strictly two-dimensional, the
dividing surface does not have any volume, and its internal structure
is not con\-sidered. While this simplifies the theoretical framework,
it neglects physical pheno\-mena which are important for understanding
fluid interfaces.
Since van der Waals \cite{VanderWaals73},
it has been understood that such a purely empirical description can
benefit from a theory of the fluid interface as a continuous
region connecting two phases.

Thermodynamically, the internal structure of the interface, such as
its thickness, can be considered by generalized versions of the Gibbs
approach, e.g.\ as devised by Guggenheim \cite{Guggenheim40} or from
more recent work \cite{FM09, LD10}.
Furthermore, investigations based on statistical mechanics can provide
a more detailed insight by describing the thermodynamics of interfaces
in terms of their molecular structure \cite{HAB76, XZDM12}. In
particular, density functional theory (DFT) in combination with
molecular equations of state was found to be a
viable approach for interfacial properties of pure fluids \cite{JDC07,
Gross09} as well as mixtures \cite{KE02, JDC07}. In combination with
simple expressions for the free energy,
DFT yields
analytical results such as the well-known approximation of the density
profile by a hyperbolic tangent \cite{Felderhof70}.

Molecular dynamics (MD) simulation, on the other hand, is based on the
equations of motion from classical mechanics. While it is
computationally more expensive, systems containing up to trillions of
molecules can today be simulated on supercomputers, employing
numerically convenient pair potentials \newydd{\cite{GK08, EHBBHHKVHHBGNBB13}}. With
relatively few model parameters, which can be adjusted to experimental
data, molecular pair potentials are highly reliable for extrapolating
and predicting a wide variety of fluid properties
consistently \newydd{\cite{UNRAL07, EVH08a}}. Both static and dynamic properties can be
computed by MD simulation \newydd{\cite{BN03, AVT09, GHV12}}, for bulk phases as well as for
heterogeneous systems \newydd{\cite{VKFH06, Mueller13}}. Even heat and mass transfer at
fluid interfaces is well accessible to molecular
dynamics \newydd{\cite{SGTB08, LVH14}}.

In a homogeneous bulk fluid, the long-range part of the force field acting on a single molecule
averages out beyond a certain cutoff radius $r_\mathrm{c}$, and straightforward mean-field approximations
can be applied to compute the long-range contribution to the energy and the pressure \cite{AT87}.
\newydd{For simulations in the canonical ensemble,} these corrections can
be treated statically for the Lennard-Jones potential, and even for dipolar molecules \cite{SFN91}, i.e.\
they have to be computed only once and do not change over time.
However, molecular simulation of heterogeneous systems is more challenging,
since the approximations behind the most straightforward techniques for
homogeneous systems, e.g.\ the reaction field method \cite{Onsager36}, break down in an anisotropic environment.

At a vapour-liquid interface, a volume integral over a
short-range interaction such as dispersion, which decays
with $r_{ij}^{-6}$ in terms of the intermolecular distance $r_{ij}$, can
yield a significant contribution, of the order of $r_\mathrm{c}^{-3}$, to the potential
energy as well as the surface tension \cite{Janecek06}. Various algorithms have been
devised to compute such effects efficiently and in a scalable way \cite{TSBI14, WRVHH14},
facilitating the massively-parallel MD simulation of heterogeneous systems with
large numbers of particles \cite{AFHLBDHKGHIHPPS13, IMHKI13}.

On the molecular level, the surface tension $\gamma$ can be considered in
different ways, based on mechanical and thermodynamic approaches. Thermodynamically,
the surface tension is defined by the free energy change related to
a differential variation of the surface area.
Such differential excess free energies can be determined by test-area simulation \cite{GJBM05, GM12},
whereas approaches based on grand-canonical sampling yield the absolute excess
free energy associated with the interface \newydd{\cite{Binder81, Binder82, SVWZB09, DB11b, TOBVB12}}.

Mechanically, an interfacial tension causes a local stress, i.e.\ a negative pressure, which
acts in the direction tangential to the
interface. For the vapour-liquid surface tension at curved interfaces, mechanical and thermodynamic methods lead to
contradicting results \newydd{\cite{SMMMJ10, MJ12, TOBVB12}}, and thermodynamic statements
cannot be based on the mechanically defined value of $\gamma$ directly.
In case of planar fluid interfaces, however,
the mechanical and thermodynamic approaches are rigorously equivalent, and
the mechanical approach, which is employed here, can be straightforwardly implemented
in terms of the intermolecular virial \cite{SM91}. If periodic boundary conditions are employed and the canonical ensemble
is simulated, the surface tension is immediately related to the deviation between the
normal and tangential components of the pressure tensor.

Accurate molecular simulation results for the surface tension require an
adequate consideration of the long-range contribution, which is sometimes nonetheless absent
from works reporting such values \cite{EVH08a, BIK13}.
Molecular models for which the surface tension has recently been evaluated reliably
include carbon dioxide \cite{GGLM08b, KRI09}, which is also considered in the present work,
water models \cite{VM07, GGLM08b}, and several other molecular fluids \cite{NWLM11, EMRV14}.
Comparing model predictions to experimental data, deviations were found
to be of the order of 10 to 20 \% for various molecular models from the literature \cite{GGLM08b, NWLM11, EMRV14} and
typically of the order of 50 \% for water models \cite{VM07}.

However, no systematic evaluation of $\gamma$ by MD simulation of an entire class of molecular models has been conducted so far. This is  the aim of the present work, focusing on a simple, but powerful class of models for real fluids from the literature. Vrabec et al.\ \cite{VSH01} and Stoll et al.\ \cite{SVH03b} developed
molecular models of the two-centre Lennard-Jones plus point
quadrupole (2CLJQ) type for 29 real compounds, including air components,
halogens, hydrocarbons, and refrigerants.
In previous work, these models were also applied successfully to
binary \cite{VHH09} and ternary mixtures \cite{HVH09}. The
vapour-liquid equilibrium (VLE) behaviour of the 2CLJQ model fluid has
been studied systematically \cite{SVHF01}, serving as the basis for a
molecular equation of state which contains an explicit contribution of
the quadrupole moment \cite{Gross05}.

A correlation for the surface tension of the 2CLJQ
model fluid from previous work \cite{WHH14} is extended by new
MD simulations in the present
work. On this foundation, the predictive capacity regarding
the surface tension of the planar vapour-liquid interface is
assessed here for these models, which were adjusted to VLE properties of the
bulk fluids only \cite{VSH01, SVH03b}, i.e.\ interfacial properties
were not taken into account for the para\-metrization.

For the present MD simulations of the surface tension, an
efficient algorithm is employed to compute the contribution of
the long-range correction \cite{WRVHH14}, combining an integration over planar
slabs \cite{Janecek06} with a centre-of-mass cutoff for multi-site
models \cite{Lustig88}. The obtained vapour-liquid surface tension is 
entirely predictive, and a comparison with experimental data can serve
to validate or improve \newydd{the molecular models}.
The surface tension predicted by these models has not been
studied previously, except for molecular nitrogen and oxygen,
where Eckelsbach et al.\ \cite{EMRV14} found a deviation of about 15 \% between model
proper\-ties and experimental data. The present work confirms this result and considers the
whole set of 2CLJQ models of real fluids systematically.

The agreement of a molecular model with\marked{ \newydd{(removed: \qq{and})}} real fluid properties, e.g.\ for the surface tension, can be improved by taking the respective experimental data explicitly into account when the model parameters are optimized.
In the literature, various optimization approaches employing a single objective function can be found \cite{EVH08a, HVMR10,HKVR10,DMVH12}. Thereby, the objective function is designed to represent the quality of several thermodynamic properties simultaneously, and specific preferences of the model developer are expressed by setting weights for these properties. To find the minimum of the objective function, gradient based algorithms can be applied, e.g. starting from a reference model from literature or based on quantum chemical calculations. The derivative of the objective function over the model parameters is evaluated and the steepest descent defines the change in the parameters.  

In the present work, a multi-criteria optimization approach is used instead to identify the Pareto set, i.e.\ the set of molecular models which cannot be altered without ranking worse according to at least one of the considered criteria. Here, several objective functions can be defined and optimized simultaneously. Since different criteria generally represent conflicting goals, it is not possible to find a molecular model leading to a minimum in all objective functions. Instead, the set of Pareto optimal molecular models (i.e.\ the Pareto set), is determined by which all possible compromises between the objective functions are accessible. Knowing the Pareto set, one can choose the model best suited for a particular application. In previous work of St\"obener et al.\ \cite{SKRHKH14}, this approach was applied to the single-centre Lennard-Jones fluid, which has two model parameters. In the present work, the four-dimensional parameter space of the 2CLJQ model is explored, yielding a comprehensive description of 
CO$_2$ 
in 
terms of three objective functions, corresponding to three thermodynamic properties: The vapour pressure, the saturated liquid density, and the surface tension.

This article is structured as follows: 
In Section \ref{sec:method}, the simulation method is briefly described.
Simulation results on the predictive power of the 2CLJQ molecular models from the literature, regarding the surface tension, are presented in Section \ref{sec:results}. Multi-criteria optimization of molecular models is discussed and applied to carbon dioxide in Section \ref{sec:optimization}, leading to the conclusion in Section \ref{sec:conclusion}.

\section{Simulation Method}
\label{sec:method}

The molecular models in the present work consist of two identical Lennard-Jones sites and a point quadrupole in the centre of mass. The Lennard-Jones potential is described by

\begin{equation}
 u_{ij}^{\rm LJ} = 4 \epsilon \left[ \left( \frac{\sigma}{r_{ij}}\right)^{12} - \left( \frac{\sigma}{r_{ij}}\right)^6\right],
\end{equation}

with the size parameter $\sigma$ and the energy parameter $\epsilon$. The quadrupole-quadrupole interaction is described by

\begin{equation}
 u_{ij}^{\rm Q} = \frac{1}{4 \pi \epsilon_0} \frac{3}{4} \frac{Q^2}{r_{ij}^5} f\left( \omega \right),
\end{equation}

where $\epsilon_0$ is the electric constant, $Q$ is the quadrupole moment of the molecules, and $f\left( \omega \right)$ is a dimensionless angle-dependent expression~\cite{GG84}.

The surface tension $\gamma$ is obtained from the difference between the normal and tangential contributions to the virial $\Pi_N$ $-$ $\Pi_T$, which is equivalent to the integral over the differential pressure $p_N$ $-$ $p_T$

\begin{equation}
 \gamma = \frac{1}{2A} \left( \Pi_N - \Pi_T \right) = \frac{1}{2} \int_{- \infty}^\infty {\rm d}y \left( p_N - p_T \right),
\label{eqn:overall-virial}
\end{equation}

where 2A denotes the area of the two dividing surfaces in the simulation volume with periodic boundary conditions~\cite{WTRH83,Janecek06} and $y$ is the direction normal to the interface.

Further technical details of the simulation method are described in the Appendix.

\section{Prediction of the surface tension of 29 real fluids by molecular simulation}
\label{sec:results}

In the following the results for the surface tension as predicted by the models of Vrabec et al.~\cite{VSH01} and Stoll et al.~\cite{SVH03b} are presented and compared to DIPPR correlations which were adjusted to experimental data\newydd{ (the employed model parameters are given in the Appendix)}. The average deviation between the DIPPR correlation and the experimental data is below 1 $\%$ for most of the fluids studied in the present work, except CO$_2$ with an average deviation of about 4 $\%$ and R115 with about 1.8 $\%$~\cite{DIPPR}.

\begin{figure}[htb]
\centering
 \includegraphics[width=7.5cm]{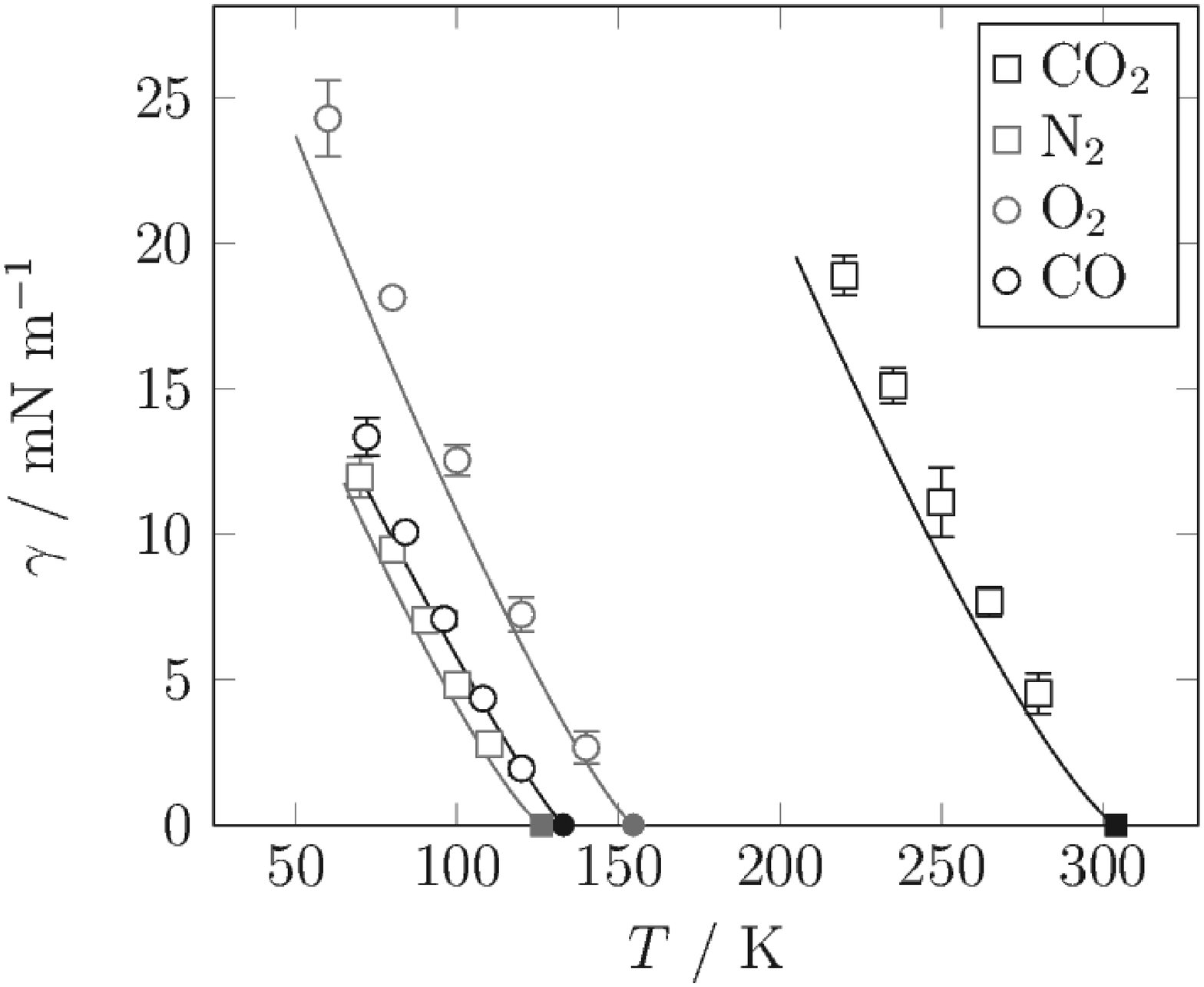}
\caption{Surface tension of air components as a function of the temperature. The open symbols are simulation results from the present work. The solid lines represent DIPPR correlations~\cite{DIPPR}, based on experimental data, and the filled symbols denote the respective critical point.} 
\label{fig:Luft}
\end{figure}

Figure \ref{fig:Luft} shows the surface tension of air components as a function of the temperature. The surface tension of N$_2$, O$_2$ and CO is overestimated by about 15 \%, while for CO$_2$ that number is 26 $\%$. The results for the surface tension of N$_2$ and O$_2$ are similar to results of Eckelsbach et al.~\cite{EMRV14}.

\begin{figure}[htb]
\centering
 \includegraphics[width=7.5cm]{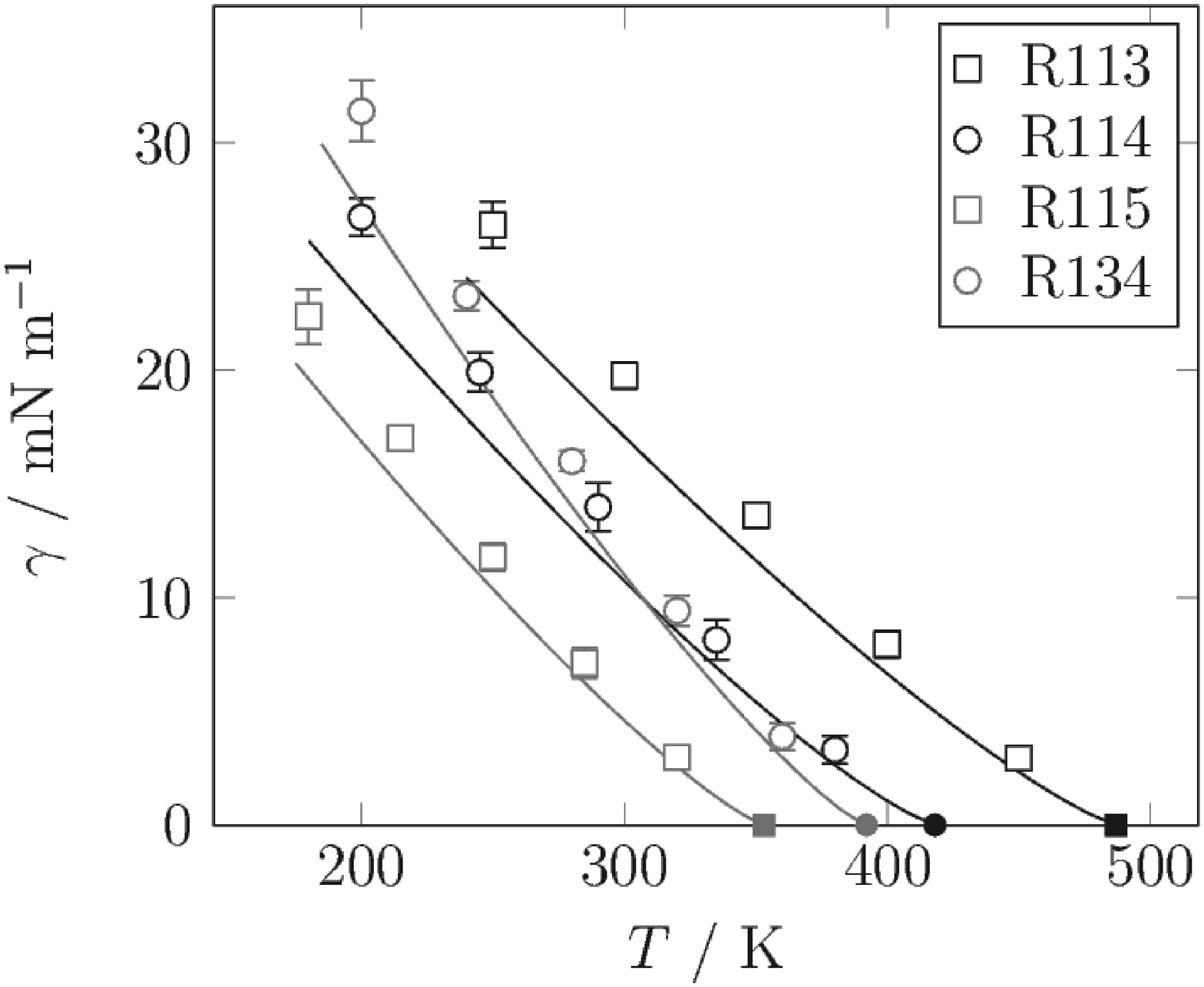}
\caption{Surface tension of various refrigerants as a function of the temperature. The open symbols are simulation results from the present work. The solid lines represent DIPPR correlations~\cite{DIPPR}, based on experimental data, and the filled symbols denote the respective critical point.} 
\label{fig:R_Mittel}
\end{figure}

Figure \ref{fig:R_Mittel} shows the surface tension of some refrigerants as a function of the temperature. The surface tension is again overestimated, but the molecular models predict the different slopes of the surface tension curve well, even though the critical temperatures of R114 and R134 are only 26 K apart.

\begin{figure}[htb]
\centering
 \includegraphics[width=7.5cm]{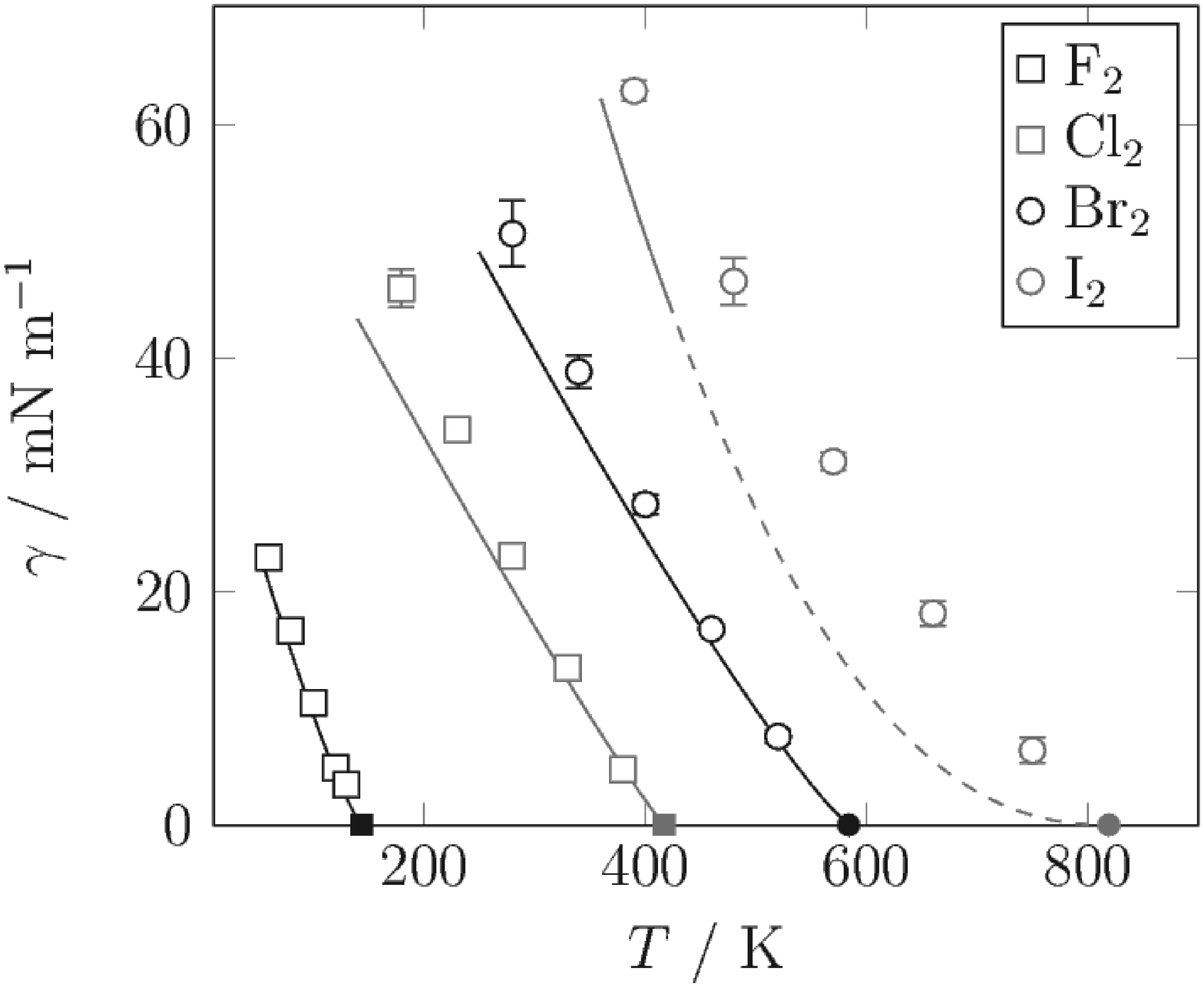}
\caption{Surface tension of halogens as a function of the temperature.  The open symbols are simulation results from the present work. The solid lines represent DIPPR correlations~\cite{DIPPR}, based on experimental data, and the filled symbols denote the respective critical point. The dashed line for I$_2$ indicates that experimental data are only available up to 425 K.} 
\label{fig:Halogene}
\end{figure}

Figure \ref{fig:Halogene} shows the surface tension of halogens as a function of the temperature. The prediction by the molecular models are about as good as in the cases discussed above except for I$_2$. Experimental data for the surface tension of I$_2$ are only available between about 390 and 425 K, while the critical point is slightly above 800 K. The extrapolation of the DIPPR correlation may be unreliable. In the temperature range where experimental data are available, the deviation between the prediction by molecular simulation and the experimental data is about 19 \%, and hence in the range as observed for the other studied systems. 

\begin{figure}[htb]
\centering
 \includegraphics[width=7.5cm]{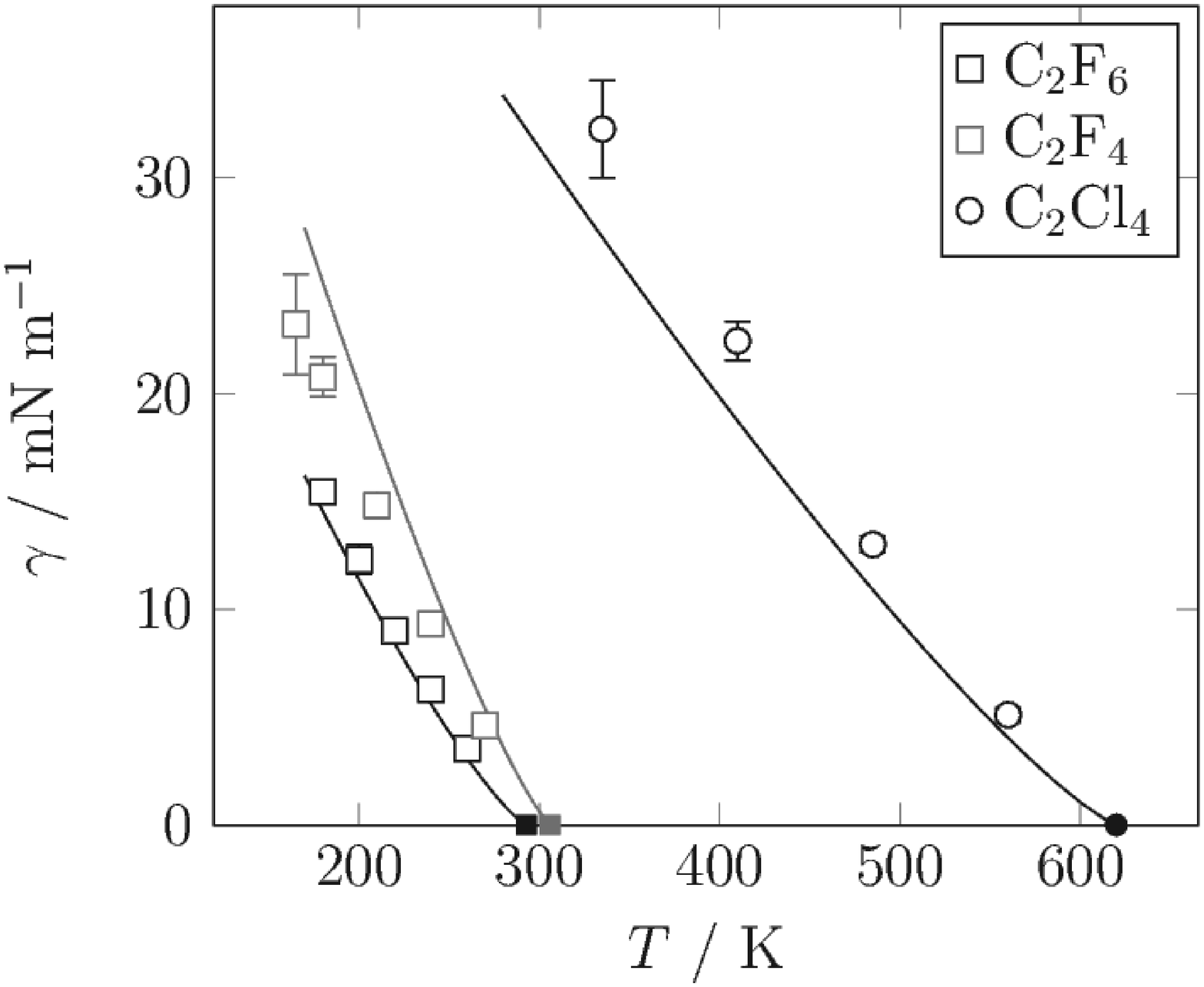}
\caption{Surface tension of halogenated carbons as a function of the temperature.  The open symbols are simulation results from the present work. The solid lines represent DIPPR correlations~\cite{DIPPR}, based on experimental data, and the filled symbols denote the respective critical point.} 
\label{fig:c2xx}
\end{figure}

Figure \ref{fig:c2xx} shows the surface tension of halogenated carbons. The molecular model for C$_2$F$_4$ is the only model that underestimates the surface tension.

The surface tension of all other compounds investigated in the present work are shown in the Appendix.

All in all the surface tension of 29 real fluids was studied in the present work. The deviation between the prediction by the molecular models of Vrabec et al.~\cite{VSH01} and Stoll et al.~\cite{SVH03b} which were not adjusted to experimental data for the surface tension is of the order of 20 \%. The surface tension is overestimated by the models in most cases. Nevertheless, considering that only data for the saturated liquid density and the vapour pressure were used for the model development, this is a good agreement.  

To increase the quality of the molecular models in terms of the surface tension, they have to be reoptimized, taking the surface tension into account.A suitable way for doing this is multi-criteria optimization. 

\section{Model Optimization}
\label{sec:optimization}

In the following a multi-criteria optimization of the molecular model of Vrabec et al.~\cite{VSH01} for CO$_2$ is discussed. Besides the saturated liquid density and the vapour pressure, which were already taken into account by Vrabec et al.~\cite{VSH01}, now also the surface tension is considered. 

Three objective functions $g_i$, depending on the molecular model parameters, are considered. Each objective function represents the relative mean deviation for one relevant property $O$, i.e. the saturated liquid density, the vapour pressure and the surface tension,

\begin{equation}
 g _i = \delta O = \sqrt{\frac{1}{N}\sum_{j=1}^N \left( \frac{O^{\rm exp}(T_j) - O^{\rm sim} (T_j,\sigma,\epsilon,L,Q}{O^{\rm exp}(T_j)}\right)^2},
\end{equation}
\marked{\newydd{(removed a superfluous comma)} }where $O^{\rm exp}$ are properties calculated by DIPPR correlations and $O^{\rm sim}$ are properties calculated by correlations to simulation data.

\newydd{The DIPPR correlations are based on the entire set of experimental data available for each fluid and deviate from the individual data points to a certain extent.} The relative mean \newydd{deviations} between the simulation data and the correlations are about 0.4 \% for the saturated liquid density, about 1.8 $\%$ for the vapour pressure~\cite{SVHF01} and about 1.9 \% for the surface tension~\cite{WHH14}. The temperature values are equidistantly spaced from the triple point temperature up to 95 $\%$ of the critical temperature in 5 K steps.

\begin{figure}[htb]
\centering
 \includegraphics[width=7.5cm]{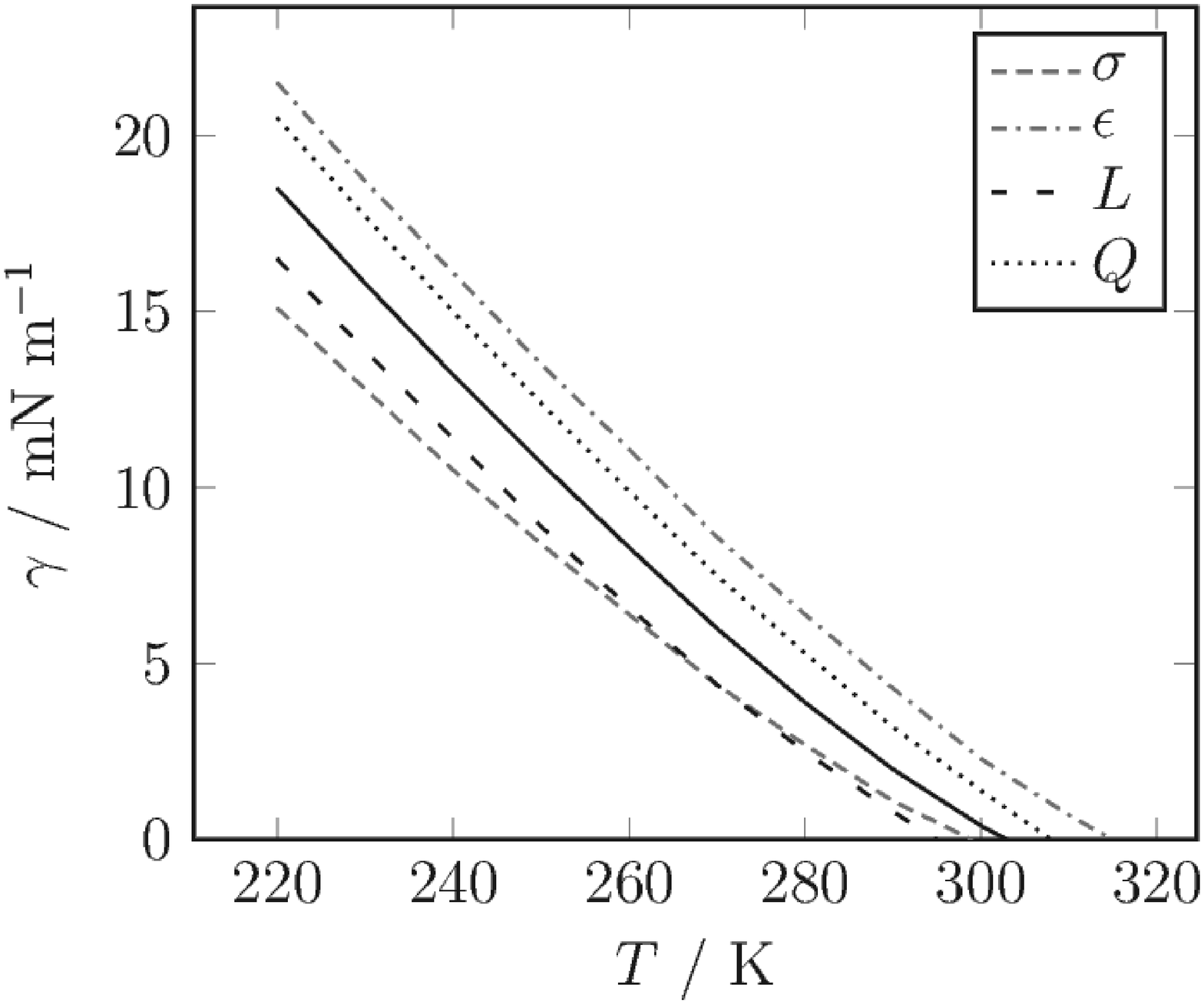}
\caption{Surface tension of CO$_2$. The solid line is the base line, representing the molecular model of Vrabec et al.~\cite{VSH01}. The dotted and dashed lines show the effect of an increase of 5 \% in the corresponding model parameter.\marked{ \newydd{(This Figure has been updated as suggested by the Referee.)}}} 
\label{fig:gammaMerker}
\end{figure}

Figure \ref{fig:gammaMerker} shows the influence of increasing one molecular model parameter by 5 \%, while the other parameters are kept constant, on the surface tension. The base line corresponds to the model parameters from Vrabec et al.~\cite{VSH01}. Increasing one energy parameter, $\epsilon$ or $Q$, increases the surface tension value, while increasing the size parameter, $\sigma$ or $L$, decreases the surface tension value. \marked{ \newydd{(Removed: \qq{$\sigma$ is the only parameter that does not change the critical temperature.})}} The corresponding phase diagram and vapour pressure curve are shown in the Appendix.

The Pareto set is obtained by brute force sampling of the parameter space. Therefore lower and upper bounds were defined for the parameters, $\sigma$, $\epsilon$, $L$, $Q$, such that the whole Pareto set is found. The sampled grid is 60 $\times$ 60 $\times$ 60 $\times$ 60 points. Based on the correlations of Stoll et al.~\cite{SVHF01} and Werth et al.~\cite{WHH14} the relative mean deviations between the simulation data and the experimental data are determined. The comparison of all molecular models generates the Pareto set.

\begin{figure}[htb]
\centering
 \includegraphics[width = 0.9 \textwidth]{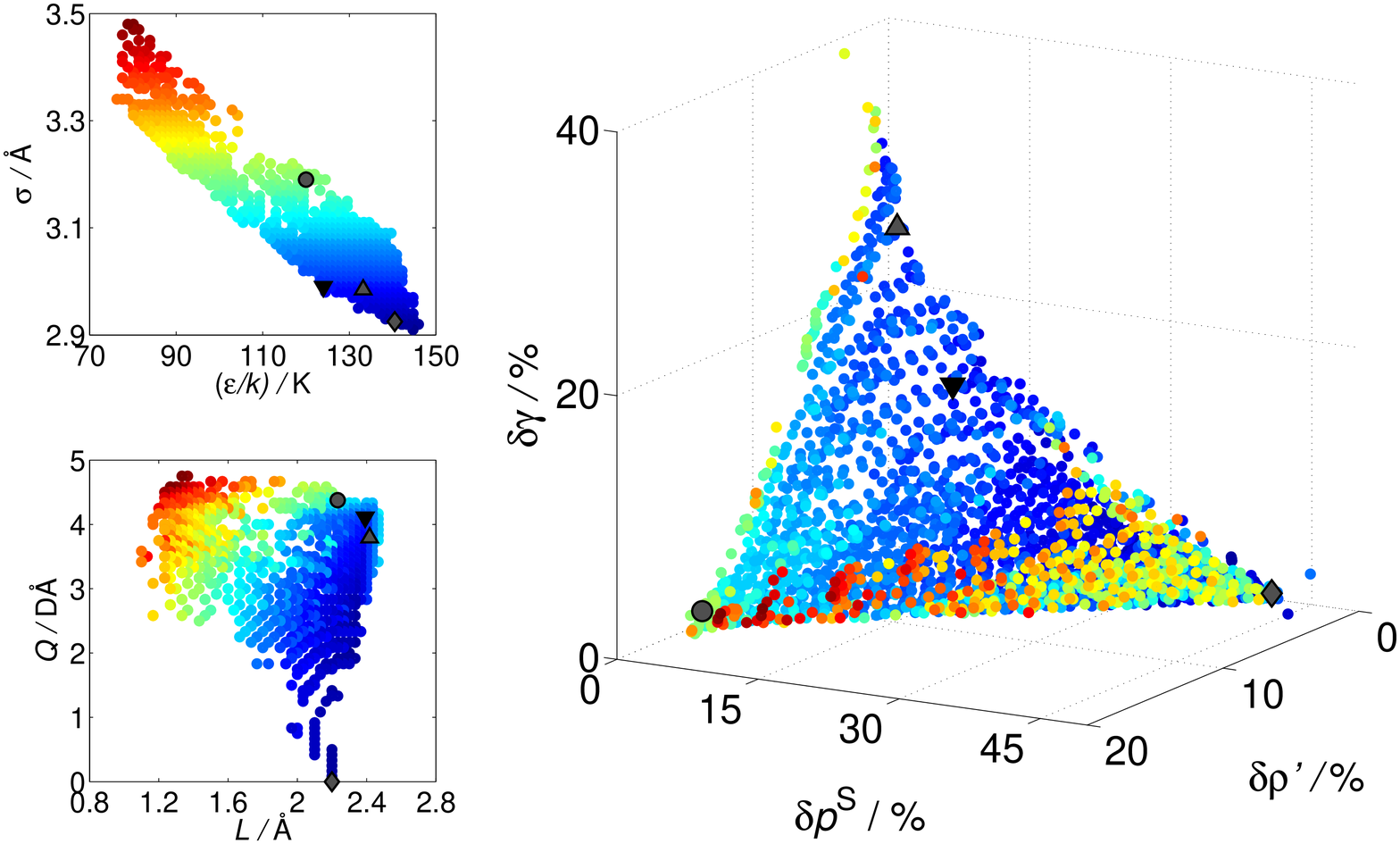}
\caption{Pareto set of the 2CLJQ molecular models for CO$_2$ in the parameter space, represented by the Lennard-Jones parameters $\sigma$ and $\epsilon$ (left top) and the model parameters $Q$ and $L$ (left bottom), and the objective space (right): deviations in the surface tension, the saturated density and the vapour pressure. The upward triangle denotes the molecular model of Vrabec et al.~\cite{VSH01}, the circle ($\gamma - p$) and the diamond ($\gamma - \rho$) denote the optimizations in two objective functions and the downward triangle denotes the new optimized molecular model ($\gamma - \rho - p$). \newydd{The colors represent the numerical value of $\mathnewydd{\sigma}$ and connect the points in the parameter and the objective space.}\marked{ \newydd{(Following the suggestion of the Referee, this Figure has been updated and improved.)}}} 
\label{fig:SigmaEpsilon}
\end{figure}

Figure \ref{fig:SigmaEpsilon} shows the Pareto set in the parameter space on the left hand side\marked{ \newydd{(removed: \qq{and the center})}} and the objective space on the right hand side, represented by the deviation in the saturated liquid density, the vapour pressure and the surface tension. From Figure \ref{fig:SigmaEpsilon} it can be seen which parameter values correspond to an optimum in two objective functions. The molecular model of Vrabec et al.~\cite{VSH01} for CO$_2$ (upward triangle in Figure \ref{fig:SigmaEpsilon}) is found to lie on the Pareto set. It represents a compromise which is excellent in the vapour pressure and the saturated liquid density, but poor in the surface tension (cf. Table \ref{tab:modelDev}). Some other compromises taken from the Pareto set are discussed in the following (parameters cf. Table \ref{tab:modelData}).
It is possible to find models which are good in the vapour pressure and the surface tension, but poor in the saturated liquid density (e.g. model $\gamma - p$ designated by a circle in Figure \ref{fig:SigmaEpsilon}) or models which are good in the saturated liquid density and the surface tension, but poor in the vapour pressure (e.g. model $\gamma - \rho$ designated by diamond in Figure \ref{fig:SigmaEpsilon}). Contrarily to the model of Vrabec et al.~\cite{VSH01} these choices are not attractive as they yield very high deviations for the quantity which is described poorly, cf. Table \ref{tab:modelDev}. Taking the model of Vrabec et al.~\cite{VSH01} as a starting point, the knowledge of the Pareto set enables finding compromises which are distinctly better in the surface tension at some expense in the quality for the saturated liquid density and the vapour pressure, (e.g. model $\gamma - \rho - p$ by downward triangle in Figure \ref{fig:SigmaEpsilon}). Note that all models discussed in the present section are optimal\marked{ \newydd{(remove: \qq{as})}} according to the definition given by Pareto.

\begin{table}
\centering
 \begin{tabular}{l||l|l|l}
	& $\delta \rho'$ / \% & $\delta p$ / \%  & $\delta \gamma$ / \%  \\
\hline
  Vrabec et al.~\cite{VSH01}	& 0.36 		& 3.68		& 26.4 \\ 
  $\gamma - p$  		& 14.4  	& 2.60		& 5.42  \\		
  $\gamma - \rho$  		& 0.77  	& 41.6		& 4.21  \\	 
  $\gamma - \rho - p$		& 0.86  	& 9.24		& 12.3  \\
 \end{tabular}
 \caption{Relative mean deviation in the saturated liquid density, the vapour pressure and the surface tension of the molecular models for CO$_2$ from Vrabec et al.~\cite{VSH01} and the optimized versions from the present work. \newline}
 \label{tab:modelDev}
\end{table}

\begin{table}
\centering
 \begin{tabular}{l||l|l|l|l}
	& $\sigma$ / \AA & $\epsilon$ / $k_{\rm B} $ & $L$ / \AA & $Q$ / D\AA \\
\hline
  Vrabec et al.~\cite{VSH01}	& 2.9847 	& 133.22	& 2.4176 &	  3.7938 \\ 
  $\gamma -  p$  		& 3.19  	& 120		& 2.233  &	  4.3766 \\		
  $\gamma - \rho$  		& 2.925  	& 140.5		& 2.144  &	   -     \\	 
  $\gamma - \rho - p$		& 2.99  	& 124		& 2.392  &	  4.1091 \\
 \end{tabular}
 \caption{Parameters of the molecular models for CO$_2$ from Vrabec et al.~\cite{VSH01} and the optimized versions from the present work.\marked{ \newydd{(The position of this table was shifted.)}}}
 \label{tab:modelData}
\end{table}
\newydd{Table \ref{tab:modelData} shows the molecular model parameters for CO$_2$ which were selected from the Pareto set of the 2CLJQ model class as described above. The molecular model $\mathnewydd{\gamma - \rho}$ does not have a quadrupole moment and the quadrupole moments of the other models are slightly larger than the value used of Vrabec et al.~\cite{VSH01}. Experimental data for the quadrupole moment are between 1.64 and 4.87 D\AA~\cite{GG84}. The experimental C=O distance is 1.15 \AA~\cite{LB61}. The deviation between this value and $\mathnewydd{L / 2}$ of the molecular models is less than 5 \%.}

More detailed information on the representation of the different thermodynamic properties by the models discussed above is available in the Appendix.\marked{ \newydd{(The position of the preceding statement has been shifted within the manuscript.)}}

\section{Conclusion}
\label{sec:conclusion}

In the present work, the ability of molecular models to predict the surface tension of real compounds was tested. 29 models of the 2CLJQ type which were parameterized using only experimental data of the saturated liquid density and the vapour pressure were used to predict the surface tension. The deviation between the prediction and the experimental data is usually of the order of 20 \% and the surface tension is overestimated. These deviations are not large considering that they refer to data along almost the entire vapour pressure curve of the studied compounds. And that both the simulation results and the experimental data are subject of errors which are of the order of 1 \% and 5 \% respectively.

Increasing the quality of the molecular models requires including the surface tension in the model optimization procedure. Here a multi-criteria optimization using a Pareto approach was used to optimize a molecular model for CO$_2$ tailored for particular applications. The Pareto approach can be generally applied to include the surface tension in the model development. A suitable multi-criteria optimization approach, based on constructing the Pareto set for the considered model class with respect to multiple thermodynamic properties, was presented here and applied to carbon dioxide. With a compromise model selected from the Pareto set, fair agreement is obtained for vapour-liquid equilibrium properties of the bulk fluid as well as the surface tension.

\smallskip

\textit{\textbf{Acknowledgment.}\/}
The authors acknowledge financial support from BMBF within the SkaSim project and from DFG within the Collaborative Research Centre MICOS (SFB 926) as well as the Reinhart Kosseleck Programme (grant HA 1993/15-1). They thank Doros Theodorou for his encouragement as well as Wolfgang Eckhardt, Manfred Heilig, Maximilian Kohns, Kai Langenbach, G\'abor Rutkai, and Jadran Vrabec for fruitful discussions. The present work was conducted under the auspices of the Boltz\-mann-Zuse Society of Computational Mo\-le\-cu\-lar Engineering (BZS), and the MD simulations were carried out on the \textit{SuperMUC} at Leibniz-Rechenzentrum Garching within the large-scale scientific computing project pr83ri.

\section*{Appendix}
\section*{Simulation Details}

The simulations were performed with the molecular dynamics code $ls1$ $MarDyn$~\cite{ls12013} in the canonical ensemble with $N$ = 16 000 particles. The parameters of the molecular models of Vrabec et al.~\cite{VSH01} and Stoll et al.~\cite{SVH03b} are given in Table \ref{tab:stollVrabecModels}.\marked{ \newydd{(The preceding sentence was previously part of the main manuscript body.)}} The equation of motion was solved by a leapfrog integrator~\cite{Fincham92} with a time step of $\Delta t$ = 1 fs. The elongation of the simulation volume normal to the interface was 80 $\sigma$ and the thickness of the liquid film in the centre of the simulation volume was 40 $\sigma$ to account for finite size effects~\cite{WLHH13}. The elongation in the other spatial directions was at least 20 $\sigma$.

\begin{table}
\centering
\resizebox{1.1 \textwidth}{!}{

 \begin{tabular}{l||l|l|l|l|l|l|l}
Name &	Formula	& CAS RN	& $\sigma$ / \AA & $\epsilon$ / $k_{\rm B} $ & $L$ / \AA & $Q$ / D\AA & Author \\
\hline
Florine		&	F$_2$		& 7782-41-4	&	2.8258	&	52.147	&	1.4129	&	0.8920 	& \cite{VSH01} \\
Chlorine	&	Cl$_2$		& 7782-50-5	&	3.4016	&	160.86	&	1.9819	&	4.2356 	& \cite{VSH01} \\
Bromine		&	Br$_2$		& 7726-95-6	&	3.5546	&	236.76	&	2.1777	&	4.8954 	& \cite{VSH01} \\
Iodine		&	I$_2$		& 7553-56-2	&	3.7200	&	371.47	&	2.6784	&	5.6556 	& \cite{VSH01} \\
Nitrogen	&	N$_2$		& 7727-37-9	&	3.3211	&	34.897	&	1.0464	&	1.4397 	& \cite{VSH01} \\
Oxygen		&	O$_2$		& 7782-44-7	&	3.1062	&	43.183	&	0.9699	&	0.8081 	& \cite{VSH01} \\
Carbon dioxide	&	CO$_2$		& 124-38-9	&	2.9847	&	133.22	&	2.4176	&	3.7938 	& \cite{VSH01} \\
Carbon disulfide&	CS$_2$		& 75-15-0	&	3.6140	&	257.68	&	2.6809	&	3.8997 	& \cite{VSH01} \\
Ethane		&	C$_2$H$_6$	& 74-84-0	&	3.4896	&	136.99	&	2.3762	&	0.8277 	& \cite{VSH01} \\
Ethylene	&	C$_2$H$_4$	& 74-85-1	&	3.7607	&	76.950	&	1.2695	&	4.3310 	& \cite{VSH01} \\
Acetylene	&	C$_2$H$_2$	& 74-86-2	&	3.5742	&	79.890	&	1.2998	&	5.0730 	& \cite{VSH01} \\
R116		&	C$_2$F$_6$	& 76-16-4	&	4.1282	&	110.19	&	2.7246	&	8.4943 	& \cite{VSH01} \\
R1114		&	C$_2$F$_4$	& 116-14-3	&	3.8611	&	106.32	&	2.2394	&	7.0332 	& \cite{VSH01} \\
R1110		&	C$_2$Cl$_4$	& 127-18-4	&	4.6758	&	211.11	&	2.6520	&	16.143 	& \cite{VSH01} \\
Propadiene	&	C$_3$H$_4$	& 463-49-0	&	3.6367	&	170.52	&	2.4958	&	5.1637 	& \cite{VSH01} \\
Propyne		&	C$_3$H$_4$	& 74-99-7	&	3.5460	&	186.43	&	2.8368	&	5.7548 	& \cite{VSH01} \\
Propylene	&	C$_3$H$_6$	& 115-07-1	&	3.8169	&	150.78	&	2.5014	&	5.9387 	& \cite{VSH01} \\
R846		&	SF$_6$		& 2551-62-4	&	3.9615	&	118.98	&	2.6375	&	8.0066 	& \cite{VSH01} \\
R14		&	CF$_4$		& 75-73-0	&	3.8812	&	59.235	&	1.3901	&	5.1763 	& \cite{VSH01} \\
R10		&	CCl$_4$		& 56-23-5	&	4.8471	&	142.14	&	1.6946	&	14.346 	& \cite{VSH01} \\
Carbon monoxide	&	CO		& 630-08-0	&	3.3344	&	36.713	&	1.1110	&	1.9170 	& \cite{SVH03b} \\
R113		&	CFCl$_2$-CF$_2$Cl& 76-13-1	&	4.5207	&	217.08	&	3.6166	&	12.984 	& \cite{SVH03b} \\
R114		&	CBrF$_2$-CBrF$_2$& 76-14-2	&	4.3772	&	183.26	&	3.5018	&	11.456 	& \cite{SVH03b} \\
R115		&	CF$_3$-CF$_2$Cl	& 76-15-3	&	4.1891	&	155.77	&	3.3513	&	9.2246 	& \cite{SVH03b} \\
R134		&	CHF$_2$-CHF$_2$	& 359-35-3	&	3.7848	&	170.46	&	3.0278	&	7.8745 	& \cite{SVH03b} \\
R30B2		&	CH$_2$Br$_2$	& 74-95-3	&	3.8683	&	274.97	&	3.0946	&	9.2682 	& \cite{SVH03b} \\
R150B2		&	CH$_2$Br-CH$_2$Br& 106-93-4	&	4.1699	&	302.33	&	3.3359	&	10.903 	& \cite{SVH03b} \\
R114B2		&	CBrF$_2$-CBrF$_2$& 124-73-2	&	4.5193	&	218.40	&	3.6154	&	12.822 	& \cite{SVH03b} \\
R1120		&	CHCl=CCl$_2$	& 79-01-6	&	4.4120	&	201.03	&	2.6357	&	13.624 	& \cite{SVH03b} \\
 \end{tabular}
 }
 \caption{Parameters of the molecular models of Vrabec et al.~\cite{VSH01} and Stoll et al.~\cite{SVH03b}.\marked{ \newydd{(This Table was previously part of the main manuscript body and has now been moved to the Appendix, following the suggestion of the Referee.)}}}
 \label{tab:stollVrabecModels}
\end{table}

The equilibration was conducted for 500 000 time steps and the production runs for 2 500 000 time steps to reduce statistical uncertainties. The statistical errors were estimated to be three times the standard deviation of five block averages, each over 500 000 time steps. The saturated densities and the vapour pressure were calculated as an average over the respective phases excluding the area close to the interface.

The cutoff radius was set to 5 $\sigma$ and a centre-of-mass cutoff scheme was employed. The Lennard-Jones interactions were corrected with a slab-based long range correction (LRC)~\cite{WRVHH14}. The quadrupole was assumed to have no preferred orientation, which yields a vanishing LRC contribution. \newydd{Following Eq.\ (\ref{eqn:overall-virial}), the surface tension was computed immediately from the deviation between the normal and tangential diagonal components of the overall pressure tensor for the whole system. Thereby, the tangential pressure $p_T$ was determined by averaging over the two tangential components of the pressure tensor.}

\section*{Additional simulation results}

\begin{figure}[htb]
\centering
 \includegraphics[width=7.5cm]{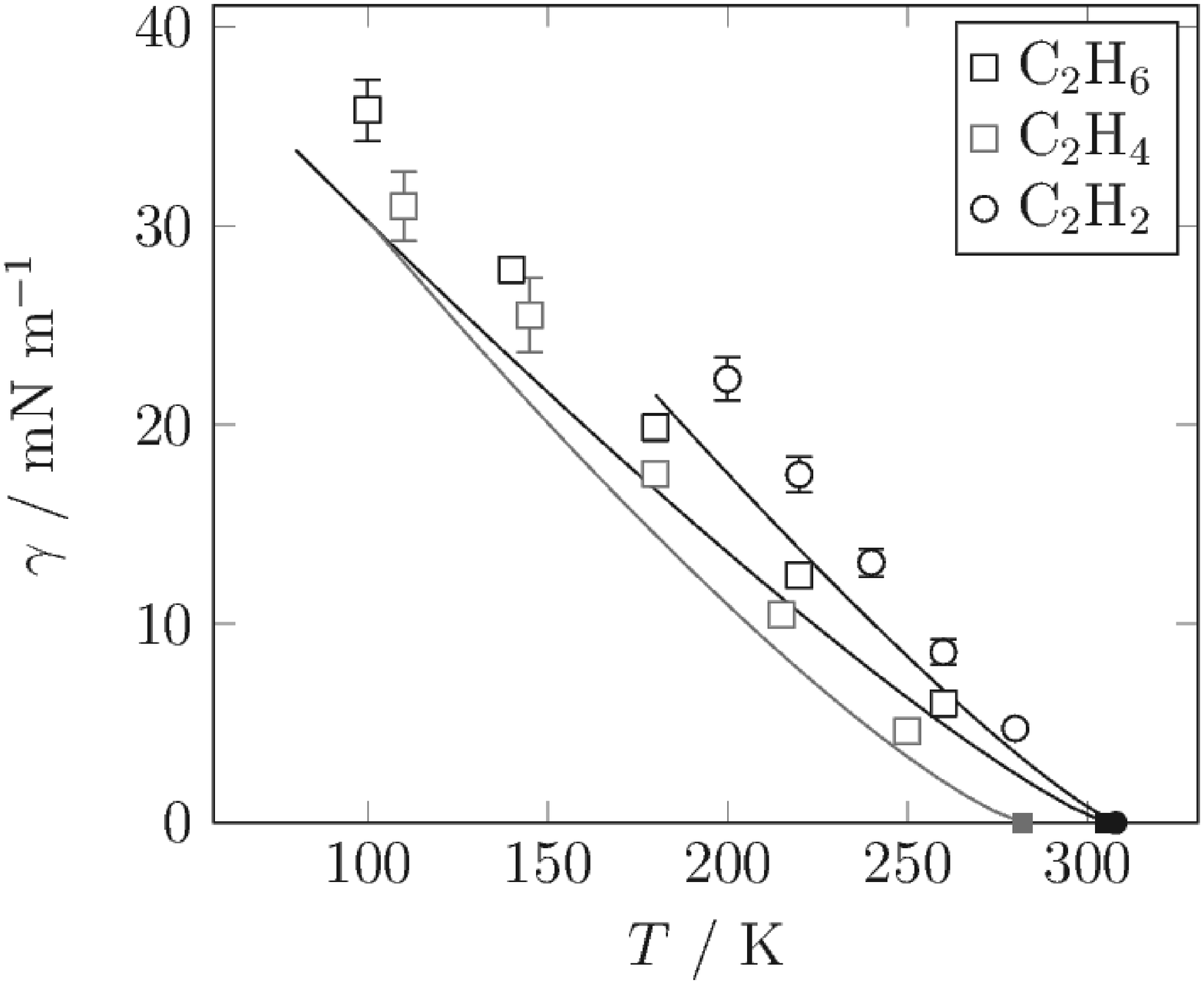}
\caption{Surface tension of hydrocarbons as a function of the temperature.  The open symbols are simulation results from the present work. The solid lines represent DIPPR correlations~\cite{DIPPR}, based on experimental data, and the filled symbols denote the respective critical point.}
\label{fig:c2hx}
\end{figure}

\begin{figure}[htb]
\centering
 \includegraphics[width=7.5cm]{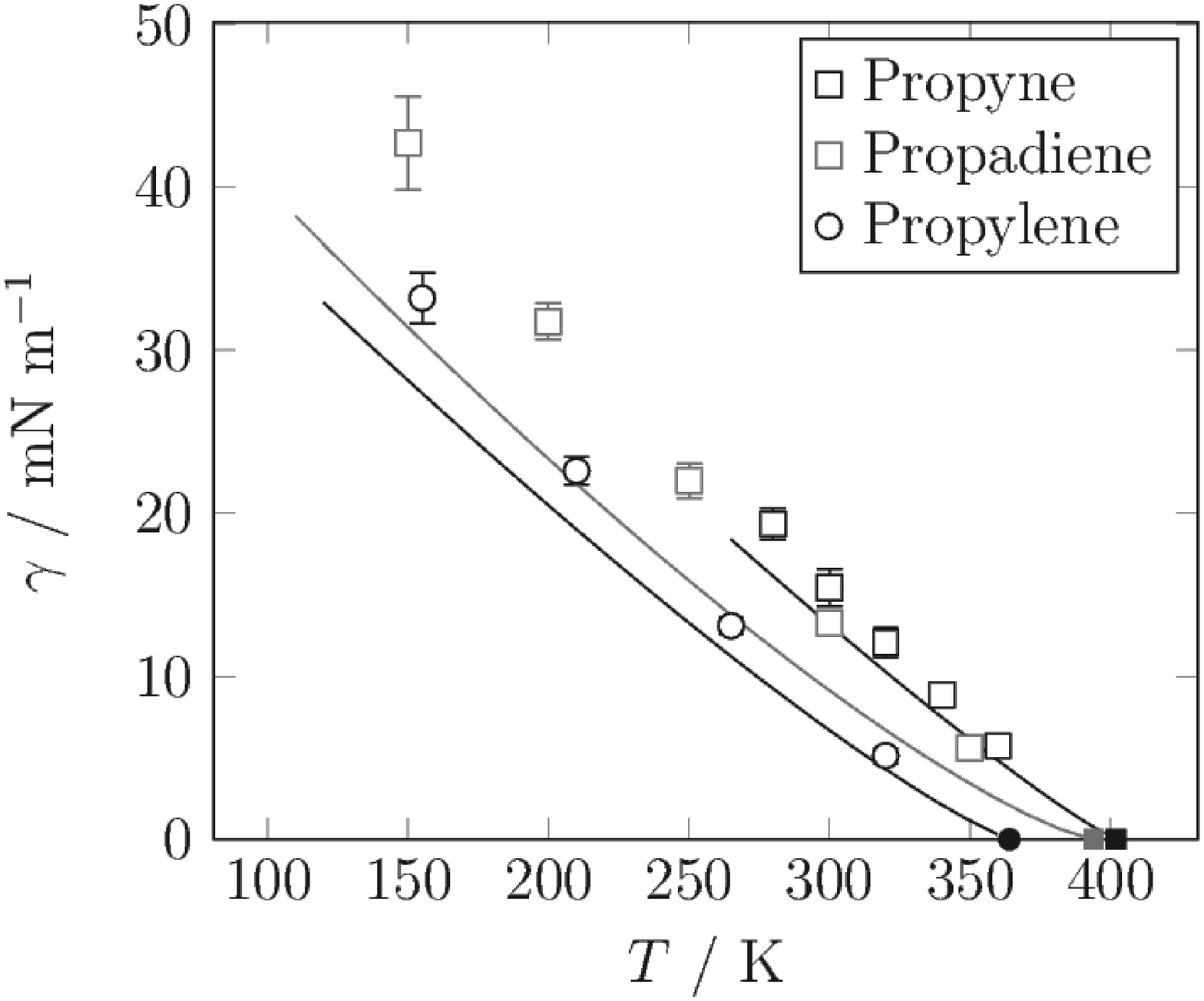}
\caption{Surface tension of hydrocarbons as a function of the temperature.  The open symbols are simulation results from the present work. The solid lines represent DIPPR correlations~\cite{DIPPR}, based on experimental data, and the filled symbols denote the respective critical point.} 
\label{fig:c3hx}
\end{figure}

\begin{figure}[htb]
\centering
 \includegraphics[width=7.5cm]{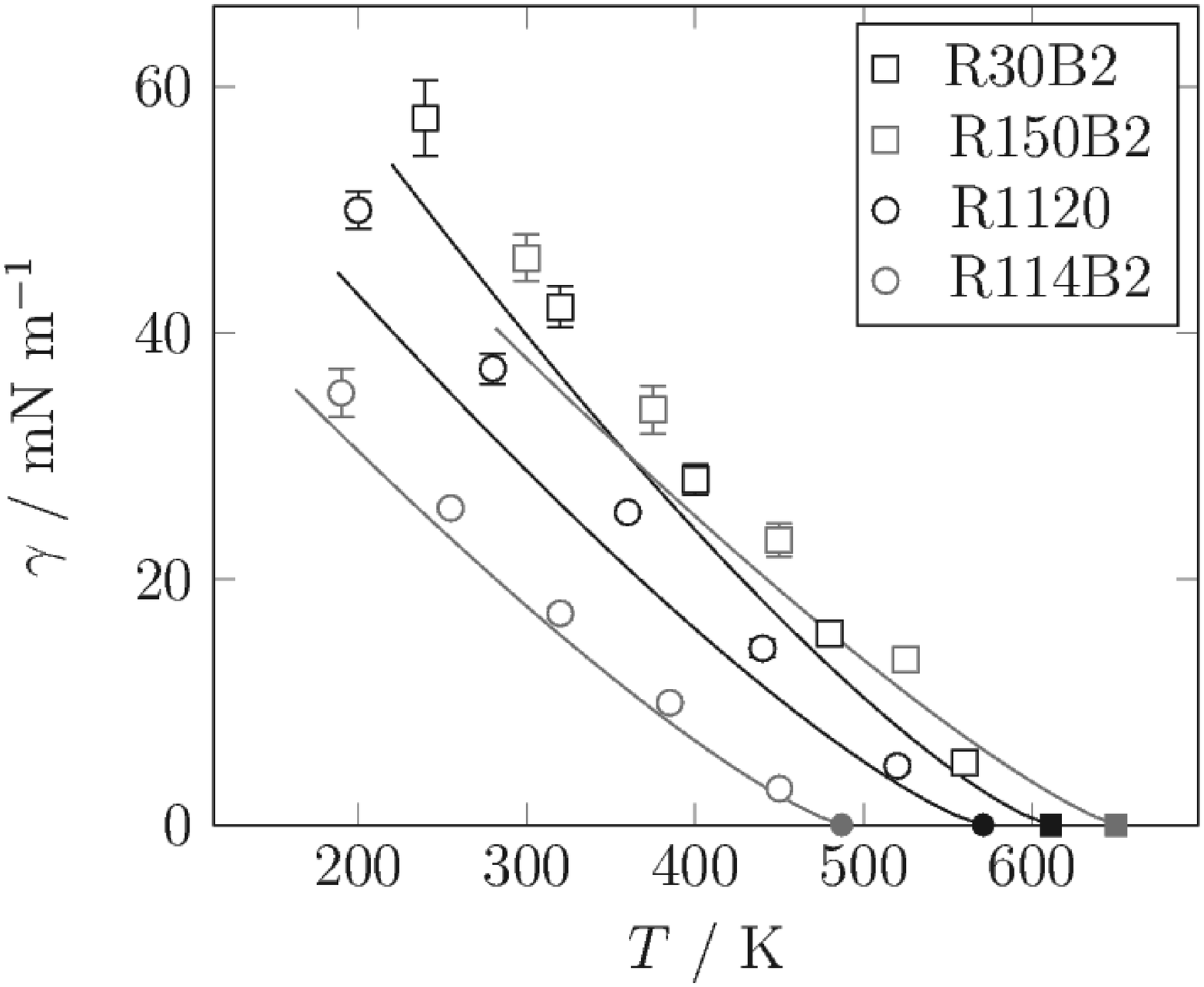}
\caption{Surface tension of refrigerants as a function of the temperature.  The open symbols are simulation results from the present work. The solid lines represent DIPPR correlations~\cite{DIPPR}, based on experimental data, and the filled symbols denote the respective critical point.} 
\label{fig:kaelte}
\end{figure}

\begin{figure}[htb]
\centering
 \includegraphics[width=7.5cm]{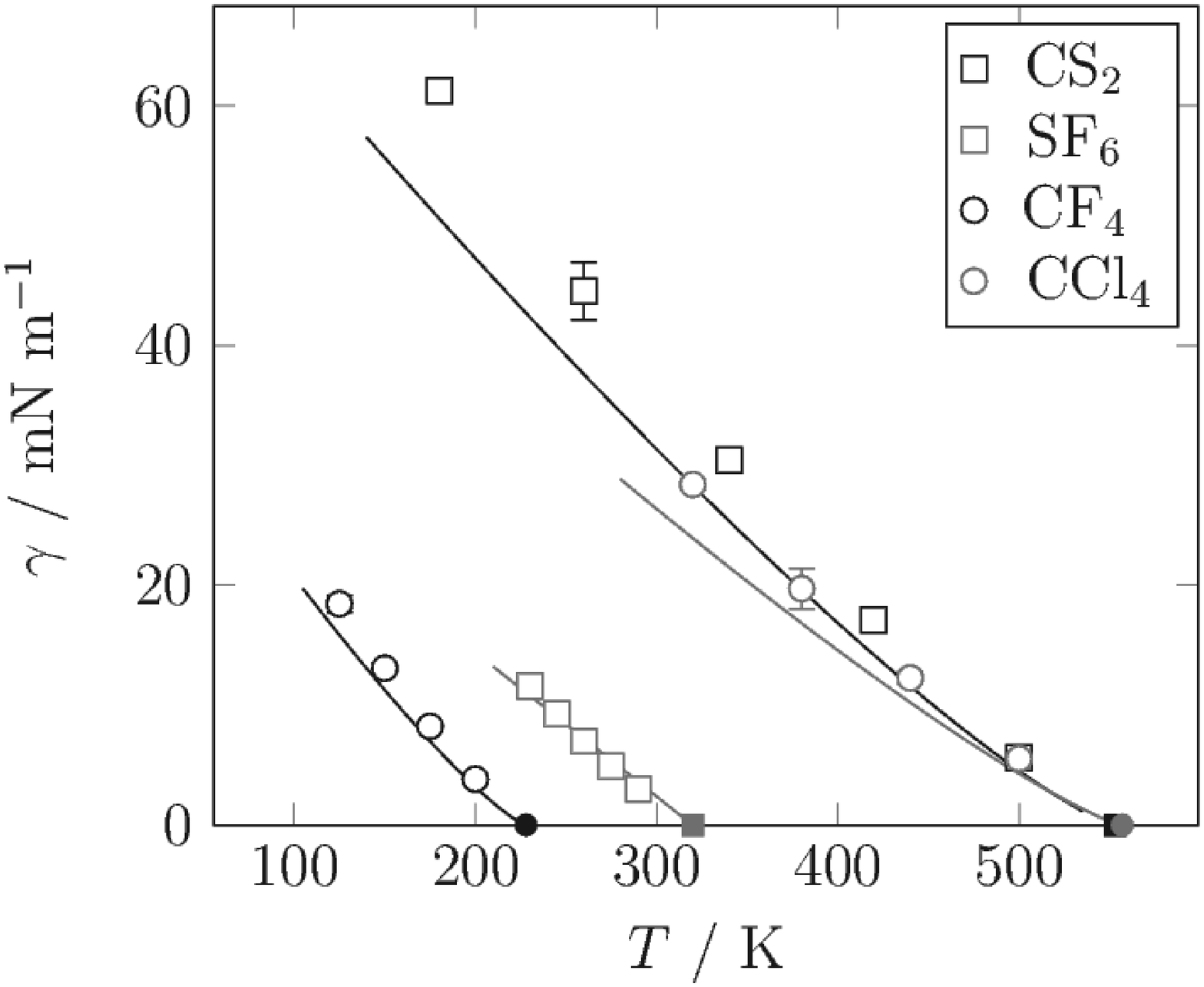}
\caption{Surface tension of various fluids as a function of the temperature.  The open symbols are simulation results from the present work. The solid lines represent DIPPR correlations~\cite{DIPPR}, based on experimental data, and the filled symbols denote the respective critical point.} 
\label{fig:rest}
\end{figure}

\begin{figure}[htb]
\centering
 \includegraphics[width=7.5cm]{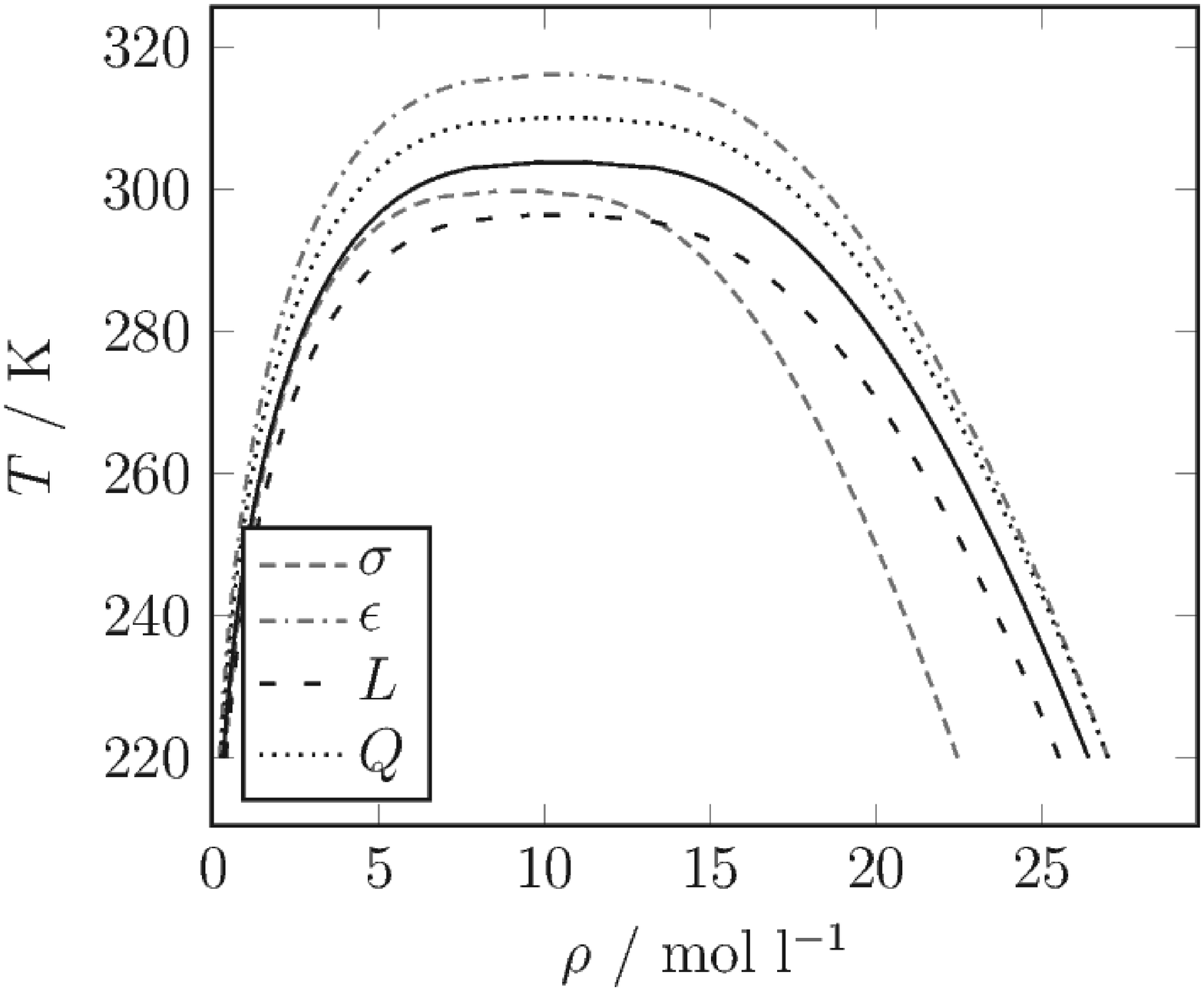}
\caption{Saturated densities of CO$_2$. The solid line is the base line, representing the molecular model of Vrabec et al.~\cite{VSH01}. The dotted and dashed lines show the effect of an increase of 5 \% in the corresponding model parameter.} 
\label{fig:rhoMerker}
\end{figure}

\begin{figure}[htb]
\centering
 \includegraphics[width=7.5cm]{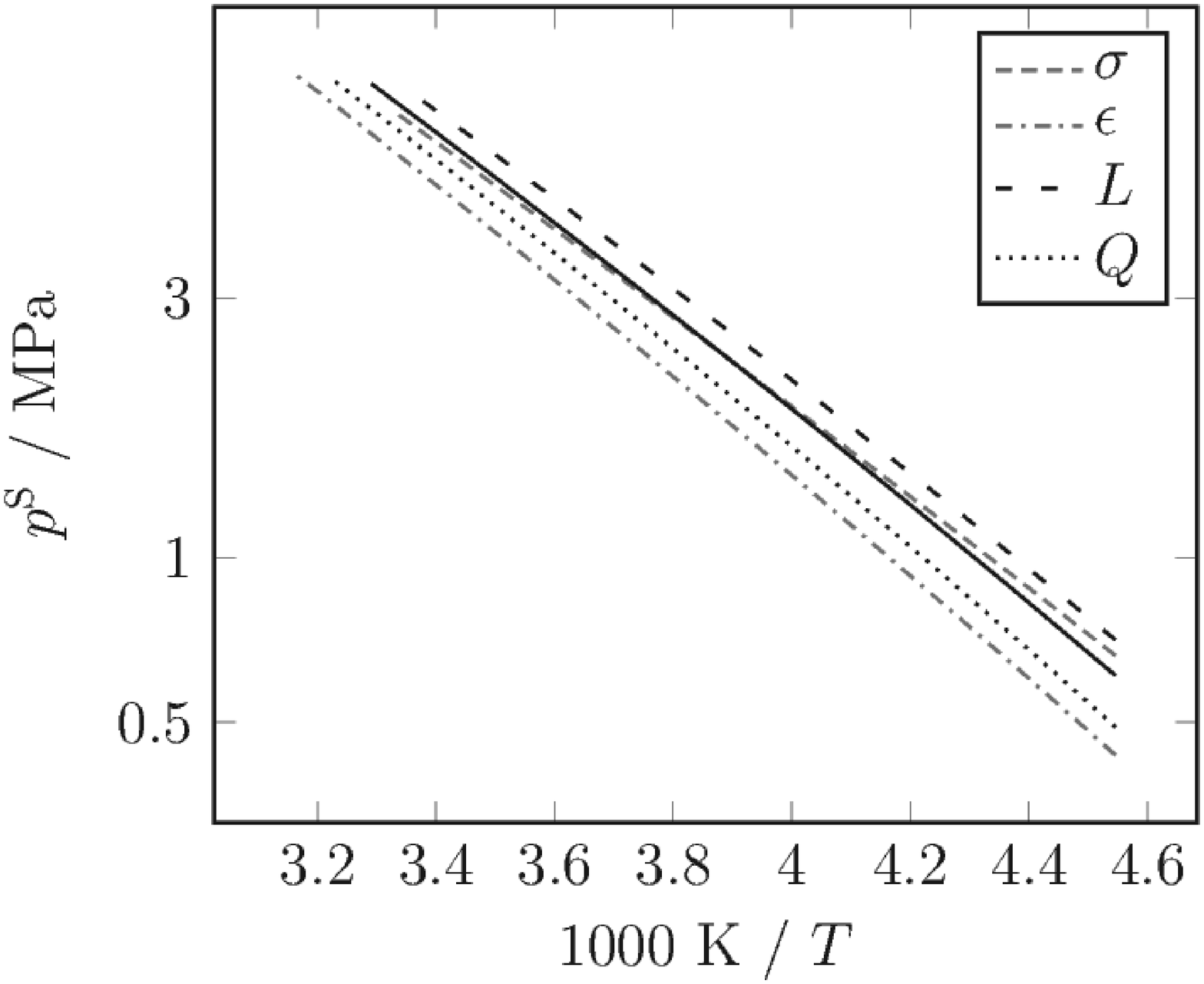}
\caption{Vapour pressure of CO$_2$. The solid line is the base line, representing the molecular model of Vrabec et al.~\cite{VSH01}. The dotted and dashed lines show the effect of an increase of 5 \% in the corresponding model parameter.} 
\label{fig:pMerker}
\end{figure}

\begin{figure}[htb]
\centering
 \includegraphics[width=7.5cm]{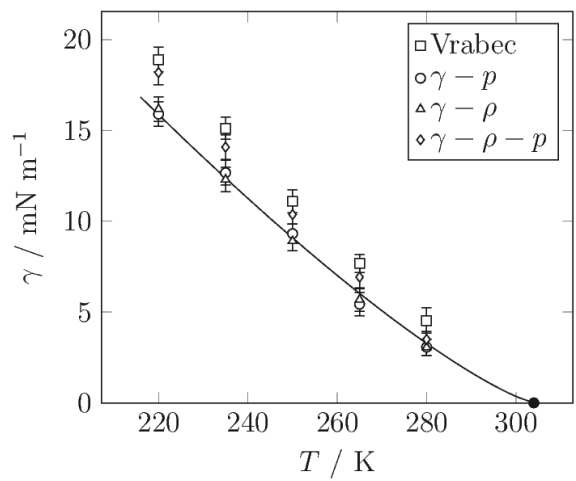}
\caption{Surface tension of CO$_2$ as a function of the temperature. Comparison between molecular models optimized to the surface tension and the vapour-pressure ($\gamma-p$), the surface tension and the saturated liquid density ($\gamma-\rho$), the optimized model ($\gamma-\rho-p$) and a previous model of Vrabec et al.~\cite{VSH01}, cf. Table \ref{tab:modelData}. The solid line represents the DIPPR correlation~\cite{DIPPR}, based on experimental data, and the filled symbol denotes the critical point.} 
\label{fig:gamma}
\end{figure}

\begin{figure}[htb]
\centering
 \includegraphics[width=7.5cm]{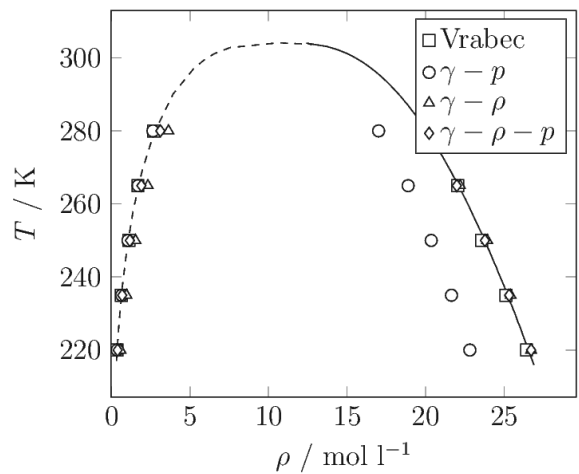}
\caption{Saturated densities of CO$_2$. Comparison between molecular models optimized to the surface tension and the vapour-pressure ($\gamma - p$), the surface tension and the saturated liquid density ($\gamma-\rho$), the optimized model ($\gamma-\rho-p$) and a previous model of Vrabec et al.~\cite{VSH01}, cf. Table \ref{tab:modelData}. The solid line represents the DIPPR correlation~\cite{DIPPR}, based on experimental data, the dashed line is based on an equation of state~\cite{SW96}. Error bars are within symbol size.} 
\label{fig:density}
\end{figure}

\begin{figure}[htb]
\centering
 \includegraphics[width=7.5cm]{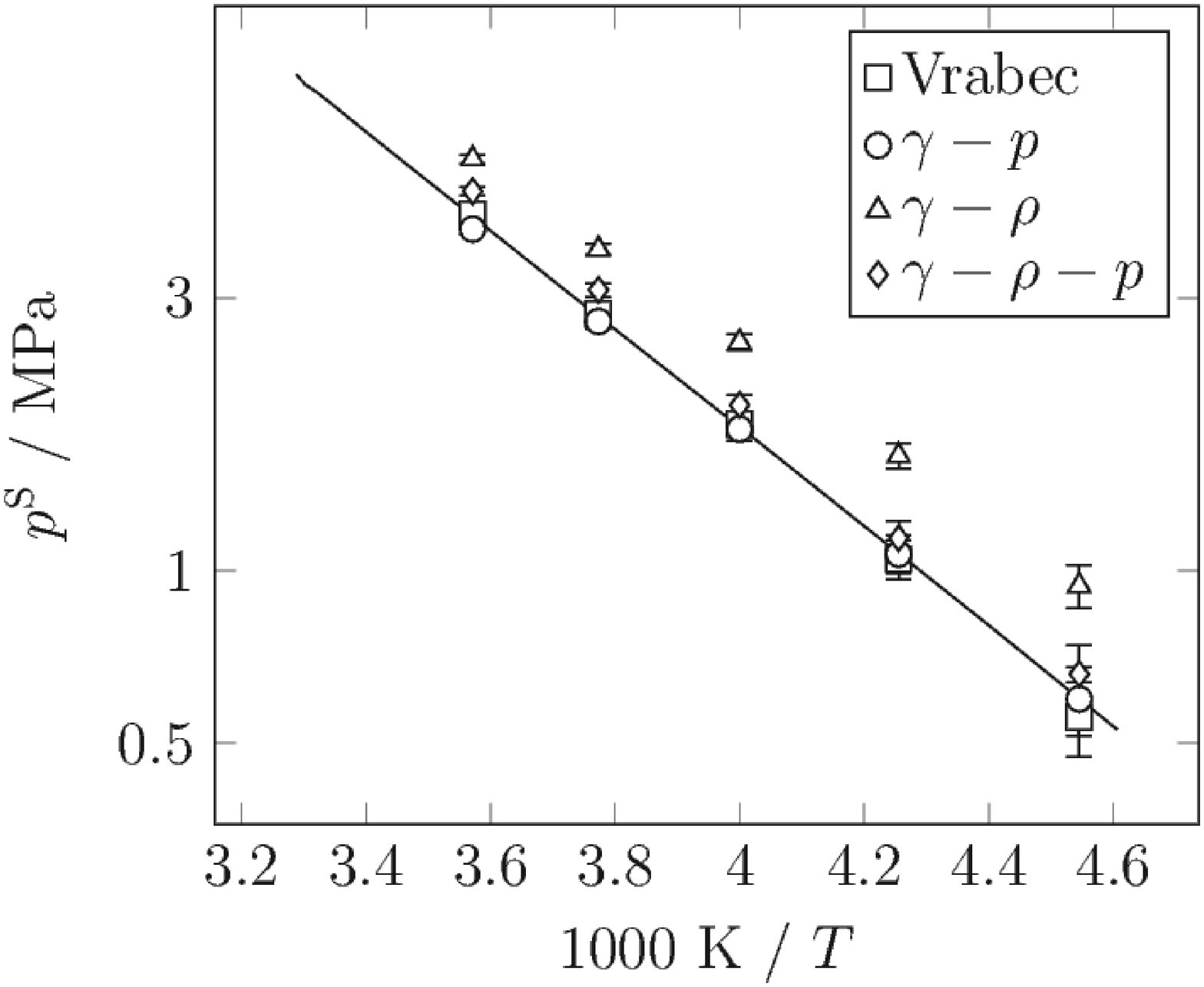}
\caption{Vapour pressure of CO$_2$ as a function of the temperature. Comparison between molecular models optimized to the surface tension and the vapour-pressure ($\gamma-p$), the surface tension and the saturated liquid density ($\gamma-\rho$), the optimized model ($\gamma-\rho-p$) and a previous model of Vrabec et al.~\cite{VSH01}, cf. Table \ref{tab:modelData}. The solid line represents the DIPPR correlation~\cite{DIPPR}, based on experimental data.} 
\label{fig:pressure}
\end{figure}

\bibliographystyle{elsarticle-num}


\end{document}